# Out of Site: Empowering a New Approach to Online Boycotts

HANLIN LI, Northwestern University, USA
BODHI ALARCON, Northwestern University, USA
SARA MILKES ESPINOSA, Northwestern University, USA
BRENT HECHT, Northwestern University, USA

106

GrabYourWallet, #boycottNRA and other online boycott campaigns have attracted substantial public interest in recent months. However, a number of significant challenges are preventing online boycotts from reaching their potential. In particular, complex webs of brands and subsidiaries can make it difficult for participants to conform to the goals of a boycott. Similarly, participants and organizers have limited visibility into a boycott's progress. This affects their ability to use sociotechnical innovations from social computing to incentivize participation. To address these challenges, this paper makes a system contribution: a new boycott tool called *Out of Site*. Out of Site uses lightweight automation to remove obstacles to successful online boycotts. We describe the design challenges associated with Out of Site and report results from two phases of deployment with the GrabYourWallet and Stop Animal Testing boycott communities. Our findings highlight the potential of *boycott-assisting technologies* and inform the design of this new class of technologies. Finally, like is the case for many systems in social computing, while we designed Out of Site for pro-social uses, there are a number of easily predictable ways in which the system can be leveraged for anti-social purposes (e.g. exacerbating filter bubble issues, empowering boycotts of businesses owned by racial, ethnic, and religious minorities). As such, we developed for this project a new, very straightforward design approach that treats preventing these anti-social uses as a top-tier design concern. This approach stands in contrast to the status quo of ignoring potential anti-social uses and/or considering them to be a secondary design priority. We discuss how our simple approach may help other research projects reduce their potential negative impacts with minimal burden.

## 1 INTRODUCTION

Online boycott campaigns have been increasing in both prominence and number. From GrabYourWallet [86] to #boycottNRA [38] to the Keurig [71] and Netflix boycotts [21], a growing number of activists are using the Internet to leverage the collective purchasing power of consumers.

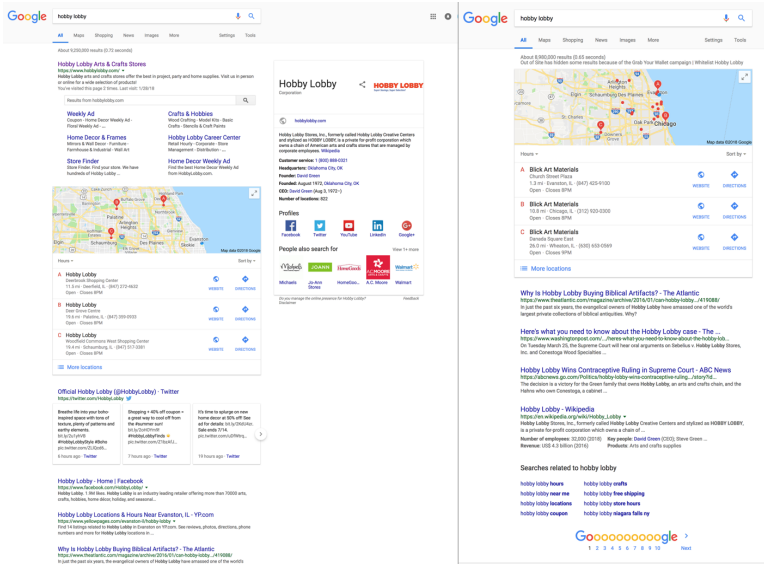

Figure 1. Left: A screenshot of unaltered Google search results for "hobby lobby". Right: The same query with Out of Site running the GrabYourWallet campaign (in "High" mode; see below). Note that almost all of the search results have changed, with the top results on the right all going to Hobby Lobby competitors. The links that made it through Out of Site's filters are discussed below.

However, despite the attention given to recent online boycott campaigns, their success has been impeded by a number of significant challenges [27,28]. For instance, it can take a great deal of individual cognitive effort to conform to the campaign's goals, with individual participants having to remember a large, potentially changing blacklist of companies before every purchase [12]. This can be even more difficult if, as is often the case, the target of a boycott owns a web of subsidiaries and a diverse array of brands. Additionally, it is often unclear to both organizers and participants how effective a boycott has been, with no way to track prevented or diverted purchases. This confusion and lack of visibility add to the already-significant sociotechnical obstacles associated with organizing any collective action campaign online, boycott or otherwise [43,48,63,67].

To address these and other challenges to online boycotts, this paper contributes a new system called *Out of Site*. A prototype of a new class of applications that we call *boycott-assisting technologies,* Out of Site uses lightweight automation to eliminate some of the obstacles to successful online boycotts. Through this automation, Out of Site is also able to track the effectiveness of a boycott and share this information with all campaign organizers and participants. This allows online boycotts to much more easily benefit from social computing's large literature on incentivizing participation in online communities [13,46,48,67,80].

Out of Site is implemented as a Chrome and Firefox web extension that allows users to join in any number of boycott campaigns. Once a user has joined a campaign, key campaign actions are automatically handled by the extension. For instance, when searching Google, Out of Site automatically hides or flags search results pointing to websites owned by a targeted company. Similarly, while shopping on Amazon, Out of Site hides or flags products manufactured by a targeted company. We designed Out of Site such that campaigns are easy to design by non-technical users, democratizing access to this new type of automation-assisted boycott.

Figure 1 shows a screenshot from a user who has joined the Out of Site campaign we implemented for *GrabYourWallet*. GrabYourWallet is a boycott community that seeks to exert economic leverage on U.S. President Donald Trump, particularly in response to the Access Hollywood tape scandal [87] and the subsequent sexual assault and harassment allegations [88]. In Figure 1's screenshot, which captures a query for "Hobby Lobby" (a company targeted by GrabYourWallet), one can see that Out of Site filtered out links to the company's website and removed the locations of its physical stores from the included map.

To understand how users interact with boycott-assisting technologies like Out of Site, we deployed Out of Site with members of two existing boycott communities: GrabYourWallet and the animal rights campaign *Stop Animal Testing*. Our deployment consisted of two phases: an 18-day study of our first version of Out of Site and a follow-up four-week study of an improved second version of the system. During the first study, we gathered log, survey, and interview data from 42 participants from these two communities. Following this first study, our research team integrated participants' feedback and additional design implications from the literature to improve Out of Site. We then deployed a second version of the system and collected more log data from the 26 first-study participants who kept the extension installed and did the same for 19 new participants that we recruited.

The results of our deployments show that Out of Site had a meaningful impact on our participants' web experiences, with hundreds of webpages affected. We also saw evidence of participants' traffic being redirected to the websites of competitors of targeted companies. More generally, participants expressed excitement about the idea of boycott-assisting technologies, although they revealed diverse preferences regarding the specific implementations of key components of these technologies (e.g. whether targeted content should be hidden or simply flagged). Interestingly, however, we also observed (and were able to interview our participants about) occasional intentional non-conformance behaviors. Specifically, some participants chose to whitelist certain targets, such as Amazon, or found workarounds to access targeted companies' websites.

While the primary contribution of this paper is the Out of Site system and its approach to online boycotts, this paper also makes a secondary contribution in the design approach we used to implement Out of Site. Computing researchers have recently come under increasing criticism for considering only the positive impacts of a paper's contribution in cases when the literature and current events make it clear that the contribution will have predictable negative impacts (e.g. [9,22,30,64,78,84]). Using a traditional design approach for Out of Site would have made this project quite liable to this critique. For instance, a system that empowers just any boycott could easily enable a powerful boycott against African American-led businesses or against businesses owned by religious minorities. Out of Site could also be used to severely bolster a user's filter bubble [77], isolating the user from any new information that might change the user's opinions about a boycott target.

To address these and other predictable anti-social uses of Out of Site, we developed a very straightforward design approach that we call *heuristic preventative design* (*HPD*). The goal of this approach is to, with minimal burden on the researcher or developer, meaningfully mitigate the negative impacts of a given system or contribution. HPD involves using the literature and current events to identify a blacklist of uses of a given system and, early in the design process, making design decisions to make those uses much more difficult. We describe how we used HPD to put obstacles in the way of the above uses of Out of Site (and others) while at the same time ensuring that Out of Site supports a diverse array of values and perspectives. We also highlight how the HPD process is sufficiently simple – it does not require specialized knowledge or extensive training – that it should be immediately accessible to a broad range of the computing community.

Below, we first present work from several disciplines that helped to motivate the idea and design of Out of Site. Next, we walk the reader through our design process. Finally, we present the findings from our deployments and discuss the implications of these findings.

## 2 RELATED WORK

In this section, we discuss the prior work that helped to motivate Out of Site and its design. This work emerges both from social computing as well as from several other domains, including consumer studies and political science.

### 2.1 Consumer Boycotts

Consumers use boycotts both to serve economic objectives, such as lowering prices, and to fulfill political and societal objectives, such as supporting fair trade [2]. In political science, boycotting is understood as a form of *political consumption*, i.e. people making purchasing decisions for political or ethical reasons [53].

Multiple studies provide evidence showing that political consumption has become a prominent form of political participation and civic engagement [1,3,76]. Indeed, political consumption research suggests that there is a large population of potential users for systems like Out of Site. For instance, in a recent U.S. survey , more than 28 percent of respondents reported engaging in at least one form of political consumption [58]. Similar studies in European countries have also observed substantial interest in political consumption, e.g. 35% of people in Sweden and Switzerland have participated in boycotts [44].

However, despite the popularity of boycotts among consumers, many barriers to successful boycotts still exist. To understand these barriers, it is useful to consider the consumer behavior literature on boycotts. Specifically, this literature has identified two types of potential boycott participants: (1) the "caring and ethical" type and (2) the "confused and uncertain" type [12]. "Confused and uncertain" consumers are those who attempt to join a boycott but are often overwhelmed with messages they perceive to be contradictory and do not receive enough guidance regarding specific actions [43]. Out of Site was designed to reduce the cognitive burden on "confused and uncertain" boycott participants by streamlining and automating boycott actions (see Section 3). Out of Site also mitigates barriers to successful boycotts for "caring and ethical" consumers. These consumers are known to struggle to integrate new information about boycott targets [10] and, as described below, Out of Site performs this integration automatically.

Aside from facilitating boycotts, Out of Site can also be understood as a system that supports a related political consumption strategy: the technique known in the political consumption literature as the "buycott" [32]. While boycott participants avoid purchasing goods, buycotts involve consumers purposefully purchasing goods from desired businesses, e.g. purchasing fair trade products [1]. While boycotting adopts a conflict-oriented strategy to punish bad companies, buycotting is centered around a cooperation-oriented strategy to reward good companies [27]. Out of Site improves consumers' exposure to non-target companies by deprioritizing boycott targets in search results, e.g. the relationship between Blick Art and Hobby Lobby in Figure 1. Through this approach, Out of Site combines some of the benefits of buycotts with those of boycotts.

Many modern political consumption campaigns are initiated by individuals online and later attract collective attention [6]. One successful example is the #deleteUber hashtag. This boycott started when a Twitter user called people's attention to an Uber promotion that took place in the context of a taxi driver strike related to changes in U.S. immigration policy [40]. The hashtag became trending in the U.S., which led Uber to make a public apology and set aside funds to support Uber drivers who were affected by the immigration policy [82]. The success of the #deleteUber campaign, however, is an outlier. Researchers have studied multiple cases of individual-initiated online boycotts and found that these grassroots boycotts rarely become effective in terms of posing economic harm to the targeted companies [72]. Out of Site was motivated in part to improve this success record, and to do so, adopting approaches from the collective action literature will be necessary. We discuss how the collective action literature helped to motivate Out of Site's design below in the sub-section that follows.

While Out of Site is by far the most advanced boycott-assisting technology of which we are aware, there are some basic tools available whose design helped to inspire Out of Site. In particular, a number of boycott campaigns have searchable databases of targeted companies and products on their websites. Out of Site effectively integrates these databases directly into participants' web browsing experiences and automates key boycott actions based on these databases. Similarly, there are a few lightweight applications that provide different interfaces to access these types of databases. One interesting such application is Buycott [89] (not to be confused with the formal term from the political consumption literature). Buycott is a mobile app that uses bar codes to query targeted product and company databases. However, Buycott is "read-only"; it does not attempt to automate any boycotting actions as in the key features of Out of Site.

## 2.2 Collective Action

Social computing researchers have examined how well-organized online collective action has led to policy changes [51], social awareness [17], successful crisis response [74], and other positive outcomes. While not directly related to boycotts, the lessons learned in the social computing collective action literature has many applications to the boycotting context. In particular, in offline boycotts campaigns and existing online boycotts campaigns, the number of participants and the aggregate economic impact are often not visible. As such, boycott participants have no way to know whether a campaign is gaining traction or has made a difference [43]. The collective action literature in social computing (and other fields) suggests that increasing the visibility of collective progress incentivizes sustainable participation and can potentially lead to larger impacts (e.g. [48,73,79]). To implement this design implication, Out of Site measures and provides real-time feedback to users about the collective progress of their boycott campaign(s). Specifically, as described in the "Design" section, the extension keeps track of how many participants have joined a boycott and the aggregate impact on participants' web experiences.

There is, however, one well-known caveat to the observed benefits of group feedback: at the beginning of a collective effort, seeing that only a few people are participating can provide a strong individual *disincentive* to participate [13,29]. This creates a well-known "chicken and egg" problem for online communities [46]. To address this problem, many design strategies have been proposed, ranging from paying professional users to attracting endorsements from prominent individuals (other strategies include creating scarce resources for early adopters, deploying bots; see Resnick et al. [46] for an overview). To address this problem, we implemented a version of the former strategy (professional users) by including boycott statistics from our pilot testers prior to the first phase of our formal deployment. In addition to practical online community design, this approach also had the benefit of ecologically validity: boycott campaigns in Out of Site will almost always be first used by individuals in the organizations that design them before they are successful with the general public.

An issue related to the "chicken and egg" problem is the "free rider problem" [63], which broadly describes when an individual gains the benefits of a group effort without making proportional individual contributions. Prior work has shown that making visible individual contributions to group progress can mitigate (and even reverse) this problem in online collective action contexts (e.g. [48,67]). Based on this literature, we designed Out of Site to prominently highlight individual boycott contributions as well as displaying group progress.

Finally, when considering collective action in online contexts, one type of technology that cannot be overlooked is social media. Extensive literature has demonstrated how social media can help catalyze successful collective action campaigns (e.g. in crowdfunding [50], social movements [11,15], crisis response [74]). Additionally, political consumption studies suggest that social media users are more likely to participate in boycotts and buycotts [5,81]. To leverage the power of social media, Out of Site contains a social sharing feature (described in more detail below) that enables users to directly engage with their social networks around their participation in Out of Site campaigns.

## 2.3 Browser Extension Research

Our decision to implement Out of Site as a browser extension was motivated in part by several recent projects that have highlighted how browser extensions can be used to alter the black-box behavior of private technologies (e.g. search algorithms) and can assist with online activism more generally. For instance, in an effort to understand how much Google depends on Wikipedia links to satisfy user information needs, McMahon et al. [52] built a browser extension that silently removed Wikipedia links from Google's search results. They deployed the extension in a small study, finding that Google search performance dropped substantially in many cases when Google could not surface Wikipedia links. This research highlighted for us the power of altering the search experience and helped to motivate the related functionality in Out of Site, although Out of Site allows campaigns to customize the search experience in a much more extensive fashion.

Browser extensions' advantages are not limited to altering search experiences; other studies have implemented browser extensions to address other power imbalances in social computing systems [37,39,55]. Turkopticon [39], built on top of Amazon Mechanical Turk, is an activist system that helps workers to publicize and evaluate requesters, a function that AMT does not natively afford. In a similar vein, Howe and Nissenbaum built an extension that allows users to directly act against online advertisers by obfuscating clicks on ads [37]. The extension, AdNauseam, extends the idea of ad blockers and simulates random clicks on the blocked ads to confuse trackers. These systems demonstrate the potential of browser extensions in empowering users to contest powerful entities, an idea that is central to boycotts.

Our decision to use a browser extension-based approach was most directly motivated by GrabYourWallet's launch of its own very lightweight browser extension approximately one year ago. The GrabYourWallet extension, which was a popular request from GrabYourWallet participants [86], has a single, simple function: it alerts the user when the user has gone to a website of a company that is on the GrabYourWallet target list. Out of Site can be viewed as a generalizable, much more powerful version of the GrabYourWallet extension. While Out of Site also can notify users when they attempt to visit a targeted website (we intentionally subsumed the functionality of the GrabYourWallet extension), it also does the difficult and critical work of customizing Google and Amazon search results pages (e.g. to benefit the millions of people who purchase through Amazon rather than going directly to company websites), works on a product-by-product basis rather than blocking entire websites, and has the extensive list of additional functionality outlined in the Design section below. Moreover, Out of Site additionally adopts design implications from the literature mentioned above (e.g. political consumption, social computing) to better facilitate collective action from interested participants. Our hope was that by subsuming the functionality of the GrabYourWallet extension while building a qualitatively more powerful system, we could build on the enthusiasm surrounding the extension to attract participation.

## 3 DESIGN

The primary contribution of this paper is a system and, as such, the primary methodological challenge of the paper was the design and implementation of the system. In this section, we outline our design process and provide details about how our implementation was motivated by related work. We also discuss how we developed and utilized *heuristic preventative design* (HPD), our lightweight design approach that integrates the mitigation of some negative uses of a technology directly into the design process. Finally, as Out of Site is a system that has been publicly deployed and thus has required frequent iterative design improvements, we also highlight the many more minor changes we have made to the system directly in response to user feedback.

### 3.1 Design Objectives

After studying the online boycott ecosystem and the existing literature, we developed three design objectives for Out of Site. The first two objectives directly address major problems in existing approaches to online boycotts, as discussed in the Related Work section above. The third objective emerged out of increasing calls for the computing community to pay more attention to the potential negative uses of the technologies it develops (e.g. [22,34,64,78,84]). More specifically, our three objectives are as follows:

1. **Conformance to Boycott's Goals:** Help people individually conform to the goals of a boycott by offloading to an automated system the logic of figuring out which items are boycotted and acting on these items (as discussed in Section 2.1 above).
2. **Collective Action:** Help people both act collectively and track the progress of a boycott (as discussed in Section 2.2 above)
3. **Avoiding Negative Impacts:** Achieve goals #1 and #2 while minimizing significant and predictable negative impacts.

Below, we organize our discussion of our design decisions using the framework provided by these three objectives. We then close by discussing iterative improvements to Out of Site that were motivated by user feedback rather than the literature.

## 3.2 Design Approach

### 3.2.1 Conformance to Boycott's Goals

As discussed above, a major challenge faced by boycott campaigns is the tremendous cognitive effort associated with successfully participating in a campaign [12,57]. This is a problem regardless of whether boycotters fall in the "confused and uncertain" category or the "caring and ethical" category (see Section 2.1). Boycott participants in the former category struggle to keep track of current targets and to recollect all of these targets' various subsidiaries and brands. In this sense, Out of Site needs to onboard this cognitive effort by identifying targets accurately and correctly in real time, and then help these participants take action on this knowledge. Out of Site also benefits consumers in the latter category, who face challenges in integrating new information at the point of every purchase [43,56]. As such, Out of Site needs to closely and timely integrate any target changes into the system.

In standard human-computer interaction terms, these issues can be understood broadly as placing excessive burden on users' recall capacity, and we know that computing technology excels at reducing recall burden [18]. As such, this major challenge to successful boycotts seemed particularly well-suited to address using computing technology. The decision to implement Out of Site as a browser extension emerged directly from this design goal, along with the prior work in browser extensions discussed above. Browser extensions can observe and change virtually any web experience (e.g. [42,52,55]), and thus can easily monitor, for instance, when a user is on a shopping website or is using Google to search for a product. They can then intervene accordingly, requiring no user recall.

This event-driven behavior, however, only assists the user in remembering to act on the boycott when they are engaging in commercial behavior. To address the recall issues associated with keeping an up-to-date copy of all brands and subsidiaries owned by all targeted companies in a given boycott, we implemented Out of Site with a server-side backend that keeps track of this information and can be updated at any time by campaign organizers. The backend is implemented as a straightforward set of keywords and web domains. These keywords and domains are currently specified manually by the campaign organizer (although we are planning a system that would automatically generate these with minimal human input). For instance, one of GrabYourWallet's target companies is the G-III Apparel Company. For this company, the set of keywords include "Calvin Klein" and "Tommy Hilfiger" and the set of domains include "tommy.com" and "calvinklein.us".

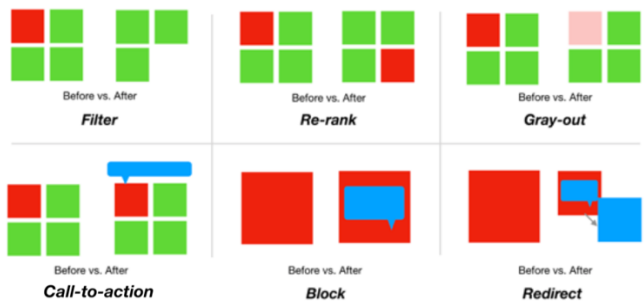

Figure 2. A conceptual diagram that depicts Out of Site's six *intervention types*. Red blocks indicate targeted content; green blocks indicate "safe" content; blue blocks indicate calls-to-action and information from campaigns.

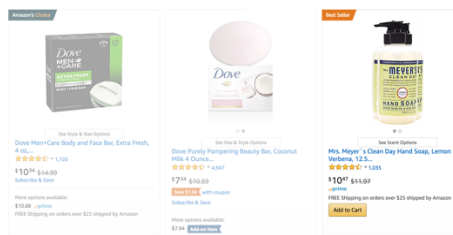

Figure 3. Demonstration of the *gray-out* intervention on Amazon search results.

Out of Site uses the lists of keywords and domains as the primary inputs to its six *intervention types*: *filter*, *rerank*, *gray-out*, *call-for-action*, *block*, and *redirect*. These intervention types are visualized in Figure 2. The *filter* intervention type removes DOM elements whose content contains any of the keywords in a campaign dataset or, in the Google case, contains links to any of the domains (*filter* is the intervention type featured in Figure 1). *Filter* requires non-trivial custom engineering to provide a good user experience on a given website, and as such, support is limited to Amazon and Google (although expanding it to sites like Wal-Mart, eBay, and Bing would simply require following the same development process we used for Amazon and Google). The main engineering challenge comes from special cases, e.g. Google's Local Search results; these require unique treatment relative to other types of DOM elements.

The *rerank* intervention type works on search results output by Amazon and allows campaign organizers to place a targeted search element at the bottom of a ranked list on a webpage rather than filtering it out entirely. Similar to *filter*, this is an intervention type that can be extended to other websites with search engines, but it does require some custom engineering.

The *gray-out* intervention type works quite similarly to *filter*, but instead of removing the targeted DOM element, it places a semi-transparent box over the element (see Figure 3). This intervention type also shows a message indicating that the element contains a link or product that is targeted by a given campaign. Our implementation of gray-out is supported for both Amazon and Google.

The *call-to-action* intervention type builds on top of *gray-out*, but instead of modifying the DOM element, the extension injects a campaign's call-to-action (e.g. sending an email to the company's PR team) around the element. This intervention aims to help boycotts that encourage participants to be vocal and directly communicate with targeted companies. For instance, for the GrabYourWallet campaign, when this intervention type is used on Google, the following moderately-sized text appears above a targeted DOM element: "{Company} is targeted by the campaign GrabYourWallet." It then asks the user to "consider contacting the company" to express dissatisfaction with company behavior. *Call-to-action* also provides a link to an e-mail pre-addressed to the company's e-mail address (as suggested by the GrabYourWallet campaign). The user only needs to press "send" in their e-mail client to complete the action.

The *block* intervention type is the most straightforward: it simply blocks users from going to a specific targeted domain, providing a message that the domain is inaccessible due to the Out of Site campaign (and provides the campaign's call-to-action). The *redirect* intervention type is similar to block, but instead of blocking the site, it redirects the user to the boycott campaign homepage after showing a message. In both these intervention types, users' visits to a targeted domain are interrupted only once within an hour. After the interruption, users can access the targeted domain without any interruption for one hour if they choose.

*3.2.2 Collective Action*

As described above in Related Work, Out of Site was designed from the ground up to incorporate several key design implications from the social computing literature on collective action. Here, we provide more details about our approach.

Communicating group progress and individual contributions are crucial for collective action success (e.g. [48,67,73,79]). Out of Site provides two types of feedback about group progress: the number of current campaign participants and the potential total impact of the campaign. Calculating the number of participants is a simple server-side measurement, and this information is displayed in the main drop-down menu of Out of Site (see Figure 4). Calculating the impact is more difficult and varies across types of campaigns. For instance, GrabYourWallet calls for participants to eschew shopping at the targeted companies. As such, for GrabYourWallet (and campaigns with similar goals), the extension highlights in its main drop-down interface how many visits to targeted websites have been blocked and how many search results to targeted websites have been altered (Figure 4). The goal of Stop Animal Testing is to avoid buying specific products. As such, for this campaign (and campaigns with similar goals), the extension displays the total number of products that have been hidden on Amazon (Figure 4). Because prior work has shown that highlighting individual contributions to group progress will increase participation in group activities (e.g. [48,67]), Out of Site also displays each user's contributions to group progress for both types of campaigns.

A new social sharing feature was implemented for the second phase of our deployment. This feature was motivated by the literature on social media and online collective action (e.g. [4,72]), and was also developed in response to user feedback that encouraged leveraging social media. By clicking on a button in the Out of Site interface, a pre-filled template message is presented that users can edit or post as-is to Twitter or to a user-selected group of e-mail addresses (implementing support for other social networks would be relatively straightforward).

Determining the best design for our template message was the subject of some literature review. In both collective action and personal informatics, researchers found that sharing active status reports are better received by audiences [19,21]. Relatedly, in existing studies about collective action on social media, hashtags have been identified as important in building and facilitating online communities [11,15]. Wikipedia researchers have also found that messages that highlight specific actions more successfully increased contributors' attention to a desired task [80]. As such, our template message includes (1) personal contributions to a campaign, (2) appropriate hashtags, and (3) a specific call for followers to join the Out of Site boycott campaign. For example, a GrabYourWallet user's message might appear as follows: e.g. "I boycotted 47 websites to support #GrabYourWallet using Out of Site (a Chrome extension). Join me now: http://bit.ly/2IkmxCq. Read about the campaign: http://grabyourwallet.org."

*3.2.3 Avoiding Negative Impacts*

Our review of the literature and current events made it clear to us early in the design process that there are a number of obvious ways that Out of Site could be used by nefarious or ill-intentioned actors to create negative impacts. These predictable negative impacts include the creation of severe filter bubbles and the use of Out of Site's boycott-assisting power by hate groups or extremists, e.g. neo-Nazis wanting to boycott all companies with Jewish and African-American CEOs.

The default approach in computing research – as was recently highlighted in a proposal from the ACM Future of Computing Academy [34] – is to consider the prevention of negative impacts like those above to be out of scope for a technical paper. That is, under the computing research status quo, the responsibility of the computing researcher who develops a new technology is the development of the technology, not mitigating the predictable negative impacts of that technology. For instance, those who develop generative models for audio and video are not expected to find ways to mitigate the disruptions to democracy that may be caused by these models [83].

With Out of Site, we wanted to try to do something different. Specifically, we sought to find a tractable design approach by which we could treat preventing predictable negative impacts as a first-order design concern rather than as an afterthought. Because of the urgent need for such an approach in a wide variety of computing research projects [14,34,78,84], we also sought to identify a generalizable process rather than one specific to this project.

As Hecht and colleagues and Parikh note [34,64], one reason why computing researchers fail to engage with the negative impacts of their research is that the effort associated with making a technical contribution in computing is already very extensive. Computing researchers have also argued (e.g. [9,14,64]) that computer scientists are usually not trained in the diverse array of social science methods and critical theory necessary for a more complex engagement with these negative impacts (e.g. [23,33,54]). For this reason, our goal was to develop a low-cost, very straightforward design approach that can address a non-trivial subset of negative impacts. While quite incomplete, such a design approach would still be a substantial improvement over current practice and would be much more easily adoptable by a wide range of computing researchers. This is similar to the logic that motivates the extremely well-known user interface evaluation technique called heuristic evaluation [62]. Heuristic evaluation gives designers a portion of the benefits of costly user studies, but is much quicker, simpler, cheaper, and more straightforward. This makes it accessible to a much wider group of computing professionals and much wider set of computing projects.

The design approach we developed we call *heuristic preventative design* (HPD), and it was inspired both by the pragmatism of heuristic evaluation and a debate that has occurred in the literature on *value-sensitive design* (e.g. [14,24,37]). With respect to the latter, an early view of value-sensitive design was based on an assumption of universal values between designers and nearly all users, with designers encouraged to incorporate these values into systems [25,26]. Le Dantec et. al. later advocated for a different understanding of value-sensitive design, one that problematized the notion of most values being universal and instead advocating that we design our systems to conform to the values of their users [47]. HPD leans heavily on the latter interpretation, but includes a critical, heuristic evaluation-like component of the former.

Like more recent interpretations of value-sensitive design, HPD asks designers to support and enable a wide variety of user values. For Out of Site, this means making it possible for a diverse set of boycott organizers with a diverse set of values to gain the benefits of the system (e.g. politically progressive or conservative boycotts). The heuristic prevention emerges in the form a specific, small set of blacklisted uses of the system that is developed by the research team. This blacklist amounts to a heuristic for the negative impacts of the system. The definition of "negative" here comes from the research team, as per the traditional understanding of value-sensitive design. The team then modifies the system design to specifically prevent the blacklisted uses.

A key component of HPD is the method one uses to develop the set of blacklisted uses. As per the pragmatism of heuristic evaluation, we suggest (and took) the following lightweight approach: following the completion of the literature review process, the research team had a meeting to brainstorm a list of potential misuses of Out of Site that could be predicted by the literature review. Given the prominence of the negative impacts of computing systems in current events (including those of social computing systems), the team also engaged with relevant recent news stories as well. Finally, motivated by the even more pragmatic approach of guerrilla usability testing [60], we also discussed the topic of potential negative uses of Out of Site with several colleagues. Within a very short period of time, we had developed a list of four potential negative uses of our system that could be easily predicted from prior work and current events. These uses and how we sought to prevent them are discussed in more detail below. Moving forward, researchers may also want to consult Ethical OS [85] (and any projects like it that emerge) when developing their blacklists. Ethical OS, which was released immediately before the final publication of this paper, is an encouraging development for HPD. It consists of a straightforward checklist intended for product teams that enumerates broad families of potentially harmful uses of a variety of computing technologies. We discuss ways that Ethical OS can be integrated into research HPD processes in the future and means by which the scientific community can help initiatives like Ethical OS in Section 5.

As discussed earlier, HPD partly relies on an older understanding of value-sensitive design in which there is some universal value set that researchers can use to make decisions about what is "use" and what is "misuse", who is a "nefarious" actor and who is a "virtuous" actor, what is a "negative" impact and what is a "positive" impact. While it is an (intentionally) imperfect assumption, the assumption of HPD is that there is a large set of uses of technology that would be considered misuse by enormous segments of the population. Examples prominent in current events include disruptions of democracy, technology addiction, and undetected and severe violations of privacy. While some may object to designing technology to prevent these broad uses, most would not. The same is true, for instance, regarding the neo-Nazi uses of Out of Site. In other words, HPD favors pragmatism over perfection in this respect. HPD is not the only approach to utilize this large-majoritarian approach; ACM leveraged a similar approach when it developed its new ethics guidelines [90].

Ultimately, however, if a researcher or user disagrees with a blacklist choice, she or he can develop a similar technology that supports these uses. We expect that if there is a reasonable argument for the mistaken inclusion of a use on a blacklist, this would be a good motivation for a new and successful research project. However, if no such argument exists, time, money, and other factors (e.g. social costs as discussed by the FCA proposal [34]) will act as significant barriers to this type of activity. These barriers could perhaps be strengthened even further through the use of intellectual property law – e.g. patents – which may give researchers and designers the legal power to exclude others from the use or reimplementation of an invention for a period of time. Under current practices, these barriers to blacklisted uses do not exist, and, as a result, computing researchers are effectively subsidizing these uses.

We emphasize that HPD with its small set of blacklisted uses is not intended to replace a more rigorous consideration of negative impacts that engages much more seriously with the social sciences and with critical theory. Instead, HPD is intended merely to replace the default status quo: *doing nothing*. Our expectation is that the HPD approach can be a stepping stone to more advanced, more complete approaches that require more training. For instance, action research [33], participatory design [54], and ethos building [23] may be appropriate more advanced approaches for projects in which HPD is used. However, as can be seen with our implementation of HPD for Out of Site (described immediately below), even this stepping stone can result in real and concrete changes to a technology.

**Implementation of Heuristic Preventative Design (HPD) in Out of Site:** As discussed above, HPD can be broken down into three key steps:

(1) Ensuring that a system is adaptable to a diverse set of user values.
(2) Building a blacklist of system uses that is motivated by the literature and current events.
(3) Designing of the system to prevent the uses on the blacklist.

Our approach to implementing the first step was to ensure that boycott campaigns are very easy to create and highly adaptable to boycott organizer preferences. Organizers can choose arbitrary keywords and domains and can decide what happens to DOM elements and sites that match those keywords and domains on a customizable basis. Indeed, our boycott campaigns are implemented as simple JSON objects that describe the various intervention types that are desired and for which websites. No campaign-specific code is included in the extension itself.

For instance, an organizer seeking to boycott Vista Outdoors (whose brands were recently dropped by REI because they also own a gun manufacturer) could simply enumerate the keywords associated with the campaign (e.g. "camelbak" and "giro", two brands owned by Vista) and the domains targeted by the campaign (e.g. "camelbak.com"). They would then select the interventions they wanted, e.g. on Google, use *filter*; on Amazon, use *re-rank*. Campaign organizers can also define the meaning of the "strength" levels (see below). All of these settings could easily be enumerated using a simple web wizard, for which we have built a prototype. However, we stopped development of that prototype for the reasons described below and we designed our two campaigns for our deployments using the raw JSON.

The process we used to develop the blacklist for Out of Site was exactly the very lightweight process we described above. Following the completion of this process, we had developed a list of four blacklisted uses. We describe these uses in detail next. We couple our description of each blacklisted use with our implementation of the third step in HPD (altering the design of the system for that specific blacklisted use).

*Blacklisted Use #1 – Enhanced Filter Bubble*: While there is some debate in the literature about where and when filter bubbles exist (e.g. [59]), most scholars agree on the negative impacts of filter bubbles that do exist (e.g. [16,65,68,77]). This is problematic for Out of Site, as one could easily leverage Out of Site as a way to block out all undesirable information about a topic, instantly inducing a sort of "filter bubble on steroids". For instance, an Out of Site a campaign could easily be built to remove all information from search results about oil companies, including removing important news stories about these companies. Similarly, one could use the infrastructure of Out of Site to remove all information from search results about a political party one does not support (e.g. a supporter of the U.S. Democratic party could block out all search results that mentioned a member of the Republican party in a campaign with the appropriate keywords).

We made specific design choices to prevent this "filter bubble on steroids" use of Out of Site. In particular, the simplest way to implement our Google SERP filtering would certainly have been to simply treat all search results in the same way, filtering out those that include targeted domains and keywords and allowing those that do not to surface to the user. However, to prevent this blacklisted use of Out of Site, we treat search results in a more nuanced fashion and Out of Site only supports the targeting of certain types of search results. In particular, Out of Site only targets search results when those results can be assumed to have a commercial intent. For instance, we do not target Wikipedia content, news stories, or any elements that are part of Google's news carousel. This means that a user who is boycotting oil companies would not see the websites for oil companies, but would see news related to the oil companies. This is particularly important in a boycotting context: news may emerge about a boycott target that might change a participant's view of the target. This news could take much longer to reach the participant without this adaptation.

Another adaptation we made to prevent this use is the introduction of boycott "strength" levels that are customizable by participants. Out of Site allows campaign organizers to set three different levels of intervention types, and users can switch between these levels. The "High" strength level is intended to be the most invasive configuration (e.g. frequently using *filter*). "Medium" and "Low" are designed to move towards less invasive intervention types like *re-rank* and *gray-out*. The strength level adaptation allows users greater exposure to targeted information if they want it, e.g. rather than search results disappearing, they would have a call-to-action around them. Below, we see that some users in our deployment moved Out of Site from the default "High" setting to the more moderate "Medium" setting.

*Blacklisted Use #2 – Use by Nefarious Groups*: Our initial plan for Out of Site was to develop an easy-to-use wizard to help any person create a campaign and to support the distributed dissemination of campaigns between users. However, given recent studies on hate speech [35], misinformation [34], and trolling [21] (and related topics), it was clear that such an approach could be co-opted by hate groups to implement campaigns with clear negative impacts, e.g. the campaign mentioned above targeting Jews and African-Americans, and similar campaigns targeting companies owned by women.

To make this use much more difficult, Out of Site now requires all campaigns to go through server-side activation. If Out of Site were to become popular, this would allow a gatekeeper to use a public policy to determine which campaigns would be supported. When instrumenting this server-side-only approach, we identified several concrete restrictions that could be included in this policy, e.g. (1) the organizer cannot be a hate group as defined by the Southern Poverty Law Center [91], (2) the campaigns can in no way select targets defined as protected classes by U.S law (supplemented by sexual preference and gender identity) [92], and (3) the extension cannot be used by a state actor. We note that such a gatekeeping approach is not novel to Out of Site. For instance, Apple's App Store takes a similar approach, in part for similar reasons [93].

*Blacklisted Use #3 – Excessive Removal of Autonomy from Users*: One controversial impact of computing systems that has attracted substantial attention is that they these systems are taking autonomy away from humans, especially in informational contexts. Ample research has shown how search engines, Facebook, and Newsfeed have impacted users' decision making (e.g. [20,45,49]). Out of Site in the most extreme cases could exacerbate this concern. Our strength levels were designed as a protection against this concern. Additionally, as the *filter* intervention type directly removes content from users' web experiences, we insert cues in users' webpages to indicate some content has been hidden. Such cues include a short sentence stating "Out of Site has hidden some results because of the {*campaign name*} campaign" and the inclusion of the number of hidden items above the extension's icon in browser (see the top of Figure 4). Additionally, we allow users to whitelist individual targets with small in-context cues displayed on affected webpages such as "Whitelist {*company name*} | Whitelist {*another company name*}". We also allowed users to whitelist boycott targets by using the extension's detailed settings.

*Blacklisted Use #4 – Causing Undue Harm*: In Out of Site, the *call-to-action* intervention type provides easy-to-use instructions for actions users can take to support a boycott, e.g. sending an e-mail to a boycott target. Such an affordance allows boycott participants to be vocal about their opinions, but if exploited, may cause excessive harm to targets (e.g. SPAM). In our implementation, we set a maximum number of daily call-for-actions that users could execute. Additionally, some of the design implications that emerge from our user study raise critical issues related to undue harm, and we highlight these issues in the Discussion section.

### 3.3 Iterative Design

Aside from the updates of Out of Site mentioned above, we have made several additional improvements to the extension in response to user feedback and usage data. Below we describe a few of the more significant improvements (all changes listed were made in time for the second deployment phase).

*Treatment of Third-Party Commercial Content:* Initially, we considered information about companies on third-party commercial platforms to be non-commercial content, e.g. we did not *filter* or *gray-out* Yelp reviews. After all, many of the Yelp reviews could be poor, and regardless, Yelp reviews may provide new, third-party information to consumers (as is the case for news articles). However, we noticed in users' search results that the costs to boycott campaigns of this approach almost certainly outweighed the benefits. For instance, in Figure 1, in an earlier version of Out of Site, a relatively positive Yelp review of Hobby Lobby appeared prominently in the post-*filter* results. As such, Out of Site now considers all prominent third-party commercial platforms to be commercial content and treats them accordingly (e.g. Yelp.com, coupon websites, links to app stores).

*Campaign Contribution Metrics:* Originally, only when Out of Site was set to "High" were users' contributions to campaigns counted towards campaign progress metrics. However, users gave us feedback that when they use "Medium" setting, Out of Site still helps them to avoid the target, and thus "Medium"-level interventions should be counted as contributions. Based on this feedback, we updated this counting mechanism.

*Simplifying Contribution Metrics*: Some users in our first deployment phase expressed confusion about the granular campaign and individual progress metrics that the first version of Out of Site provided. To avoid this confusion, Out of Site now presents the simplified metrics shown in Figure 4.

*Installation Wizard:* Our interviews following the first deployment phase made it clear that many users were not aware of some of the features in Out of Site, nor were they aware that they could opt-out of any of the installed campaigns. As such, for the second deployment phase, we implemented an installation wizard that had users opt in to campaigns and, for each campaign, choose the "High", "Medium", or "Low" setting.

### 4. USER STUDY

To better understand how people interact with Out of Site and boycott-assisting technologies more generally, we conducted an in-the-wild user study. To do so, we released Out of Site into the Chrome and Firefox Web Stores for participants to download. We then recruited people interested in our two proof-of-concept boycott campaigns to install Out of Site. We collected survey data, interview data, and log data from participants. The study was developed in concert with the research team's local IRB and was eventually determined to be exempt due to relatively strict anonymization procedures and restraint in the data that was captured.

Below, we first provide more detail about our two proof-of-concept campaigns and why they were selected. We next discuss the context and approach of our user study, which emphasized in-the-wild, exploratory observation. Finally, we present our quantitative and qualitative results, which are organized together into four themes that were established through an affinity diagramming process.

### 4.1 Proof-of-Concept Campaigns

For our user study, we implemented Out of Site campaigns for two existing boycott communities: GrabYourWallet and Stop Animal Testing. The GrabYourWallet boycott is a grassroots effort aimed at companies that have any connection or businesses with U.S. President Donald Trump and/or his family. The boycott has a list of targeted companies on its website[1] and a team of organizers monitors the companies on this list and updates the list when necessary. The Stop Animal Testing campaign is led by People for the Ethical Treatment of Animals (PETA). This campaign highlights a list of cosmetic and household products that it suggests people avoid due to the animal testing that was used to develop these products[2]. We selected these two campaigns because they have attracted a large number of participants. This meant that we would have a large potential population of extension users.

While choosing this combination of campaigns provided the benefit of integrating with two well-established boycotts, this combination also presented a few challenges. In particular, the GrabYourWallet campaign is boycotting Amazon as a company (as well as many of the companies that sell products on Amazon.com). As such, in our first deployment phase, for users who accepted our default settings, their Amazon links in Google Search were filtered out and some of their visits to Amazon were redirected to the GrabYourWallet campaign website. This likely reduced users' visits to Amazon.com, which are already much lower than those to Google, and prevented us from gathering as much data about interaction with Amazon.com and the corresponding intervention types as we expected. In the case of our first deployment, the limited data we did gather from Amazon.com came from users who turned off the GrabYourWallet campaign, set its strength to "Medium" or "Low", went to Amazon.com more than once in an hour, or whitelisted Amazon.com. The installation wizard mitigated these issues in the second deployment, although interestingly it did not lead to a major increase in data from Amazon.

### 4.2 User Study: An Exploratory and In-the-Wild Approach

Following the deployment of Out of Site to the Chrome and Firefox stores, we sent out recruitment materials to existing online communities and message boards that are related to the GrabYourWallet and Stop Animal Testing boycotts. For instance, we advertised in Facebook groups and sub-reddits that are dedicated to women's rights issues and animal welfare. We also recruited members of our local community interested in these topics.

As noted above, this paper reports the results of two separate phases of deployment. In between the two phases, we did extensive development based on feedback and log data from participants in this first phase, as described in the Design section. We use the term "first version" and "second version" to distinguish between the versions of the extension used in each phase. Our first deployment phase had 54 installations, with 42 people using the extension more than one day. This first phase lasted three weeks and average usage time was 6.7 days (although this was substantially attenuated by users who signed up midway through the first phase, leaving less time in the phase for usage). The second phase lasted four weeks and included 26 users from the first phase who had continued using the extension (the in-between phase data was not considered), as well as 19 new users who were recruited through a new round of advertising (21 new installations; two used the extension less than one day). The average usage length in the second phase was 15.8 days. We restrict our log analyses to people who used the extension for more than one day, although we analyze the available data on less-than-one-day users as well to gain a better understanding of non-use.

---

[1] https://grabyourwallet.org
[2] https://www.peta.org/action/easy-ways-help-animals-used-killed-experiments/

Immediately after installation, users were asked to fill out an optional survey about their prior experiences supporting boycotts and other civic campaigns. The 48 users who completed the survey (36 from the first phase; 12 from the second phase) reported that, as predicted by the political consumption literature, many of them (46) had already been incorporating company ethics into their consumer behaviors and have had experiences with boycotting a variety of organizations, including the NRA (39), Uber (33), and Wal-Mart (18).

In the first phase, the extension's default settings enrolled users into both the GrabYourWallet and the Stop Animal Testing campaigns, with both campaigns' strength level set to "High". However, users were able to customize their enrollment and settings freely and could turn off one or both of the campaigns easily. In the second phase, users were walked through these key settings in an installation wizard.

The log data we collected through the extension was limited to two types: (1) how participants interacted with the browser extension itself, e.g. turning it on and off, whitelisting targets, and changing strength levels and (2) statistics from web pages that are affected by the extension. These log data were then uploaded to our database every 24 hours. In an effort to protect our users' privacy and in accordance with our IRB, our extension only collected information about visits to pages it had modified. Specifically, only altered Google SERPs, altered Amazon pages, and direct visits to targeted websites were recorded by the extension. For similar reasons, we also did not collect identifying information in the log; no experimenter is able to directly tie any specific user to their log data (although research has shown log data can be used for deanonymization with some effort [35]).

To collect qualitative feedback, we reached out to participants via email with a request to interview them during the first deployment (e-mail addresses were not tied to log data). Seven users responded to our emails and were interviewed over phone or via text messages for approximately 30 minutes on average. Interviews were open-ended to elicit both generic feedback about the boycott-assisting technologies concept and granular feedback about Out of Site's settings and features. Each interviewee was compensated with a $10 electronic gift card, which was sent out via email at the end of the interview.

Participants were permitted to delete the extension from their browsers whenever they desired. When a user uninstalled the extension, an optional exit survey was shown to elicit any final feedback that users might have.

### 4.3 Results

To understand our results, we first combined our (1) interview data (which we transcribed), (2) our survey data, and (3) written observations from an exploratory log data analysis. We then conducted a standard affinity diagraming process. Affinity diagramming is a popular approach among HCI researchers and practitioners to identify themes in heterogeneous data (e.g. [7,8,31,66]). Specifically, to execute the affinity diagramming, two members in our team created codes stemming from our three sources of data. We conducted two sessions of diagraming and iteratively refined our themes as new data became available. The final output of the affinity diagramming process was a set of four themes that cut across our quantitative log data and our qualitative survey and interview results. Data from our second deployment were combined into these four themes later to support or contrast with our previous findings.

The four specific themes are: (1) Out of Site had a meaningful impact on users' web experiences, (2) there was a tension between automating action and automating awareness, (3) participants had positive reactions to collective action features, and (4) there was some non-conformance to the campaigns' goals.

*4.3.1 Out of Site had a meaningful impact on users' web experiences*

Out of Site affected 660 and 480 web pages in the first and second deployment phases, respectively. Across both deployment phases, the vast majority of the web pages were changed because of the GrabYourWallet campaign (655/660, 480/490). This distribution is not unexpected: the Stop Animal Testing campaign was a much more targeted campaign, only affecting Amazon and only a limited number of target companies (a portion of this effect is also likely due to the interaction between the campaigns discussed above).

The GrabYourWallet campaign had a truly substantial impact on participants' Google search experiences; search engine results pages (SERPs) were the venue for the vast majority of the campaign's interventions (539 of 655 pages / 440 of 480 pages). Across both deployment phases, the *filter* intervention removed a total of 165 advertising links, 207 "knowledge graphs elements" (e.g. information boxes on the right side of search results [52]), 27 links to Twitter.com and over a thousand (1,065) standard search links. In the second deployment, Out of Site additionally filtered out 37 links to third-party commercial websites (e.g. Facebook, Trivago, coupon websites, Yelp).

GrabYourWallet participants who used the "Medium" setting also saw consequential changes to their Google SERPs, although they were of the *call-to-action* intervention type rather than *filter*. Across both deployment phases, 214 search result links (of all types) were marked with a call-to-action on a total of 164 SERPs. We also saw a few users experimented with the call-to-action links that provide users an e-mail to send to the targeted companies. We only saw four users in total used the "Low" setting in the GrabYourWallet campaign.

Content-wise, the majority of the Google links affected were those to e-commerce websites. Amazon was by far the most impacted website. However, Macy's, Wal-Mart, Bed Bath and Beyond, Papa John's and Chewy.com were also affected in decent numbers. Indeed, examining the results, it is clear that Out of Site had its greatest effect for *transactional queries* [69] (e.g. "New Balance 530", "Papa John's"). This is the type of query most associated with commercial purchases in Rose and Levinson's three-part search query schema [69], meaning that Out of Site is having the intended effect of intervening in potential commercial transactions. Correspondingly, the SERPs that had fewer affected links usually were the result of "navigational" or "informational" queries, the two other types of queries in Rose and Levinson's schema. For instance, the query "chromebook video showing green" only had an Amazon link to a Chromebook affected. Similarly, the query "6.5 us to cm" resulted in a SERP that only filtered out a link to 6pm.com.

Outside of Google, in the first deployment phase for GrabYourWallet, 116 direct visits to companies' websites were either redirected to GrabYourWallet's website (*redirect* intervention type; "High" setting) or blocked (*block* intervention type; "Medium" setting). The second phase saw a total of 40 redirected or blocked pages.

Although the Stop Animal Testing campaign did not generate a large amount of log data, we observed that 40 animal testing products were affected on Amazon.com across 13 search queries in the first deployment phase. Similarly, 40 products across 10 search queries were affected in the second phase. As an example, one query for "Skinfood fresh fruit lip" resulted in the removal of products from ChapStick and Maybelline, two brands that were targeted by the campaign.

Importantly, with respect to Out of Site's supporting of "buycotts", we observed evidence from both deployments that Out of Site diverted clicks to competitors of targeted companies. For example, in our first deployment, a user searched for "america's first civilization michael coe" and received a SERP that had Amazon.com links removed. This user then actively engaged with multiple alternative search result links, including to Barnes & Noble and abebooks.com (an online book store). Similarly, in our second deployment, a user searched for "Mihelcic and Zimmerman, 2nd edition" and received a SERP with filtered-out Amazon.com links. This user then visited a variety of online book stores such as abebook.com and wiley.com.

With respect to our interview data, participants expressed almost exclusively positive opinions of the extension's impact on their browsing experience (although this result is of course subject to observer-expectancy effects, novelty effects, and sampling bias). One interviewee (P4) tested the extension immediately after installation and was "excited to see it worked so well". Similarly, another interviewee (P6) remarked with excitement that Out of Site is a "passively active approach" that "allows people make a social impact without having to do anything". Another interviewee said of the GrabYourWallet campaign "It's difficult to constantly keep track of all the businesses you interact with that oppose your values... It looks like [your] list is (automatically) updating."

*4.3.2 Tension between automating action and automating awareness*

We saw significant evidence of users altering the strength settings (i.e.,"High", "Medium", "Low") of their campaigns. In the first phase, five participants switched the GrabYourWallet campaign to "Medium", one did so for Stop Animal Testing, and five users experimented with different strength settings and returned to the default "High" setting. In the second phase, because of our installation wizard, a number of participants (7/19) chose to use "Medium" or "Low" settings ("Medium" was far more popular).

From our interviews, it is clear that some users appreciated that the "High" setting automated their boycotting *actions*, while others felt that this setting was too invasive and instead wanted the extension to automate their *awareness*, i.e. by flagging targets when they encountered them. This latter group found the "Medium" setting to be the most effective.

P7, a member of the group who remained on the "High" setting spoke very positively of its ability to automate actions: "I signed up for PETA's mail list and followed a couple of advocates on Twitter. I usually look for information online, like looking into news articles, just Google it. I really like the Stop Animal Testing campaign... It (the extension) has been something I am looking for. It filters out things automatically."

Two of our interviewees set the extension to "Medium". One of these participants (P2) remarked, "I want to know what is blocked and when, so I don't miss anything important... the information will still be there if I need it." Seen in an HPD light, this user benefited from and appreciated the intervention we implemented to prevent the filter bubble use case. The other participant (P1) who switched the extension to the "Medium" setting did so to become more familiar with the extension before trusting it to take action on her/his behalf: "I am a tinkerer. I used [the] 'Medium' setting to get familiar with what the extension does. I need some time to go on [the] 'High' setting."

Overall, these data reveal a tension in how to implement Out of Site's boycott-assisting automation: some users appear to want actions automated (e.g. content filtered out). Others simply want the extension to automate the awareness process, helping them to understand when content from targeted companies is surfaced, but allowing them to take their own actions. For now, this suggests that Out of Site's implementation of a user-configurable setting is advisable, but there is likely more that can be done to navigate this tension. We return to this point below in Discussion.

*4.3.3 Collective action features caused excitement*

As predicted by the social computing collective action literature, participants reacted quite positively to Out of Site's collective action features. For instance, one interviewee (P6) echoed the findings in this literature that increased visibility of group progress incentivizes individual participation: "I like that information showing how many people joined you and how many products were hidden. It shows people are making progress."

Similarly, another interviewee (P1) pointed out that the visibility of campaign progress could also help the campaign achieve its action-oriented goals: "If such thing snowballs, it could make a bigger impact. The effects of economic boycotts aren't often immediately visible. This could make the impact visible to large corporations."

Furthermore, interviewees envisioned how the extension could be integrated with existing organizations and online communities to coordinate collective action: "I hope there would be a way I can communicate with others, like I can directly contact the organization if I have any questions." (P6). Similarly, other interviewees expressed the need to leverage social media to disseminate calls-to-action. As described above, this feedback was in part what motivated us to integrate social media features into the second version of Out of Site.

*4.3.4 Non-conformance was observed*

Participants in our study at times attempted to evade the automated boycotting assistance provided by Out of Site to meet an immediate need. To do so, some users employed the customization features that were provided by the extension. For instance, in our first deployment phase, out of five users who chose the "Medium" settings for the GrabYourWallet campaign, three of them changed their setting to "Medium" after failure to retrieve needed results from Google search.

One interviewee (P5) emphasized what he believed to be the necessity of having the strength features in the extension for this purpose: "I think the simplicity of this extension itself is a good tool, but sometimes 'Medium' would be better if there's only one company making a thing on amazon, or if you just need to get some Papa John's for whatever reason, especially because they have a big discount for college students."

In addition to lower strength levels, users also employed the whitelist feature as a workaround for accessing targets. 17 users leveraged this feature in total across both deployment phases, 16 of whom whitelisted Amazon (other whitelisted targets include Papa John's, chewy.com, Bed Bath and Beyond, Wal-Mart, US Bank, and Belk.com).

We also saw evidence that some of the users who uninstalled the extension (see above) did so for reasons related to non-conformance. For instance, in one of our six exit survey responses, one user mentioned that the extension blocks shopping websites and causes inconvenience: "The few times I needed to shop online I couldn't use normal sites at least not for a while." We also wondered if users who engaged with the extension for less than one day dropped out for the same reason (14 users in total) but were unable to infer more information from their log data. The majority of these users only opened the settings of the extension once and did not otherwise engage with the extension.

## 5 DISCUSSION

Our deployments provided early evidence that Out of Site's vision of boycott-assisting technology has significant potential. Below, we first detail a number of implications for the design of boycott-assisting technology that arise from our research. Next, we discuss ideas for advancing our approach to HPD. Finally, we close by highlighting several important limitations of our research.

### 5.1 Design Implications

*Customization Capabilities are Important:* Our qualitative data suggests that participants valued and benefited from the customization capabilities included in Out of Site. In particular, in our user study, we observed that users have very different preferences between automating awareness and automating actions, and customization allowed them to adapt Out of Site to their preferences. Future work could take one step further and provide users with even more choices. For example, a feature could be added that allows users to set dates and times of participation, e.g. Mondays through Fridays or 10:00-14:00 every day, which may reduce campaign drop-out rates.

*SERPs are an Effective Site of Action*: Another major insight from our user study is that the adaptation of search engine results pages (SERPs) can be a very effective mechanism for boycott campaigns. In our two short deployments, Out of Site made well over 1,000 changes to our participants' SERPs. Boycott-assisting technologies – and likely other collective action campaigns – are likely well-served by focusing development effort on contesting the information delivered in SERPs.

*Automation and Activism:* At a high level, our user study provides evidence of the promise that automation holds for boycott-assisting technologies, as well as for activist and civic technologies more generally. Although our results suggest that this automation should be paired with significant user customization capabilities, it is clear from our qualitative data that the automation in Out of Site enabled our participants to act on their values at scale in a way that they found empowering. Out of Site is part of a family of automation-assisted activist and civic technologies (e.g. ResistBot [94]), and our results suggest that this family should grow.

Such automation strategy has been widely applied to other fields such as computational journalism to monitor events and produce content [49]. In a similar vein, activists could benefit from automation technology that helps to monitor social and political issues and take simple actions (e.g. advocating on social media).

*Community Functionality is Desired*: Although Out of Site currently does not have a social aspect, multiple interviewees requested the ability to directly communicate with campaign organizers and other participants. Existing platforms that mobilize collective efforts such as change.org and gofundme.com provide a space for campaign participants to share their personal stories and motivations, as well as for campaign organizers to provide periodic feedback. It could be useful to integrate similar functionality either into Out of Site or on an associated website.

*How to Scale Up?:* As discussed above, our HPD process revealed a major tension between the capability of Out of Site to support basically any boycott by any community and our desire to ensure that Out of Site is not used by hate groups and related organizations. This resulted in us pausing development of the easy-to-use wizard that outputs the JSON object that defines each campaign. Moving forward, it would be ideal to have this wizard in place and hosted online, but to couple this with the sociotechnical development of the gatekeeping process described above. This would make Out of Site resemble Apple's App Store: like Apple's SDKs, our wizard would provide the capability to easily build a powerful tool, but we would also have a rigorous submission and approval process before the tool is launched.

Another factor to take in consideration while scaling up such boycott-assisting technology is how to deal with conflicts between campaigns, e.g. the issue related to Amazon in our deployments. One way to address these types of issues is for users to be able to rank their campaigns by priority. Out of Site could then use these ranks to help users resolve these conflicts.

*Automated Keyword/Domain Identification*: As noted above, one of few non-trivial tasks associated with building a campaign for Out of Site is identifying the set of keywords and domains associated with targeted companies' subsidiaries and brands. GrabYourWallet's official extension addresses this challenge by blocking all the domain names that are listed on their campaign website. However, we noticed that as many of the targets are conglomerates, such as Amazon, their subsidiaries sometimes are not flagged as targets in the extension.

In the process of developing Out of Site, we identified conglomerates' subsidiaries using data from Wikipedia and public records. This was not much of a burden even for GrabYourWallet, which has many targets, and many of these are conglomerates. However, we did find that the manual approach was somewhat error-prone: in our first deployment, we accidentally missed a few of Amazon's subsidiaries (e.g. IMBD and Goodreads). Although this was easily fixable as keyword and domain tracking is managed on the server side, this also highlighted for us the importance of developing an automated keyword/domain generation tool. Using such a tool, an organizer could simply add the company names of which the organizer is aware and the relevant keywords and domains for all subsidiaries and brands would be output automatically. This is likely a tractable problem given the increasing availability of semantic web data (e.g. Wikidata) that contains subsidiary and brand relationships. Indeed, this problem is currently on the development shortlist for Out of Site.

*Addressing Non-Conformance*: Above, we saw examples of participants who found ways around Out of Site's boycott guardrails to engage in activity not encouraged by the targeted boycotts. Prior work has found boycotts are most likely to succeed when purchasing the targets' goods or services is a highly visible action [27]. This visibility is much weaker in online settings, where other participants cannot see a person walk into a targeted store or walk out of a store with a targeted item. Fortunately, by mediating the online experience, Out of Site is well-situated to address this downside of online boycotts. This will have to be done with care, however. To shame individual users publicly would almost certainly result in supporting our blacklisted use related to undue harm, although in a new way (as can be predicted by the online harassment literature [4,41]). One more positive approach might be to show an anonymized, aggregated statistic in our group progress display that indicates how many non-conforming visits and purchases have occurred in total.

*Replacement Discovery*: Our interviewees reported cases in which alternative sites or products to those targeted by a boycott were not available, which is consistent with prior studies in the boycott literature [27]. If alternatives are not available, boycott participants do not have any options other than purchasing from the boycott's targets. Fortunately, some campaigns have started aggregating alternative options to recommend to their participants. For example, PETA has a large database consisting of "cruelty-free" brands that do not use any animal testing based on their research. Ethicalconsumer.org provides a list of companies' ethical rating that takes multiple factors into consideration (e.g. environment, social responsibility). Out of Site could use these databases to power a recommender system that could be integrated into Amazon and Google search results as a new intervention type.

### 5.2 Advancing Heuristic Preventative Design

While we have enumerated how we utilized HPD for our project above, it is useful to briefly consider how it might be operationalized for other projects to explore its generality. We believe that for nearly all the research domains mentioned in the FCA proposal [34], HPD could provide useful insight and likely mitigate some negative impacts. For instance, in the case of the generation of audio and video with neural networks, HPD would likely result in a blacklist that includes the use of the neural networks to make propaganda. This would then encourage the research team to find ways to build watermarking or related approaches into the core of their approach (rather than treating it as a separate problem). Similarly, a research project that advances brain-computer interfaces might generate a blacklist that includes unwanted read/write access to specific parts of the brain. They would then work to prevent that use case within the initial contribution. Finally, a research team building a tool that semi-automatically tracks food consumption and encourages healthy eating would likely want to blacklist a use in which people with eating disorders co-opt the tool to advance their disorder.

If HPD were to spread in popularity, a clear and important next step would be the creation of an aggregate list of well-motivated potential problematic uses of various types of computing innovations. As noted above, this list could derive from or build off the recently released "Ethical OS" checklist and would serve as a key input to project-specific blacklists. Such a global list would be an analogue to the standard heuristics used in heuristic evaluation [61,62]. Such a list would additionally further reduce the burden on researchers and developers employing HPD, potentially adding to HPD's broad accessibility. However, one challenge here would be adequate summarization and navigability, which already may be a risk with the Ethical OS checklist. Additionally, researchers should take care to ensure that each heuristic is well-motivated by the literature, with news stories providing the sole motivation only when literature is not (yet) available. This is one means by which researchers can support Ethical OS and similar projects.

### 5.3 Limitations

Although (and perhaps because) Out of Site advances social computing's understanding of a new problem space – automation-assisted boycotts – our research is subject to some limitations. First, as is often the case with deployed systems, Out of Site had a few bugs, especially in the first deployment phase. For instance, as noted above, for the first-phase GrabYourWallet campaign, IMBD and Goodreads were not flagged as subsidiaries of Amazon (this was fixed for the second phase). It is also worth noting that some of our search queries on Google or Amazon might be users merely experimenting with the extension. We do not expect, however, that either of these issues had a meaningful impact on our exploratory user study and its high-level observations.

Another limitation of the study is that we tracked a very small fraction of users' browser histories in an effort to protect users' privacy. As such, we were unable to identify additional websites or content that could have been targeted (e.g. it could be that supporting Wal-Mart's search function is important for, for instance, GrabYourWallet members). It is also important to note that as both proof-of-concept campaigns share somewhat similar low-level political ideologies, our user study's result might not apply to other demographics, as different ideological groups may adopt different tactics [36].

## 6 CONCLUSION

In this paper, we have described Out of Site, a *boycott-assisting technology* that automates many of the challenging aspects of implementing successful boycotts. We described the unique design approach we took with Out of Site that we call *heuristic preventative design* and reported on the use of Out of Site in two deployments with 42 users and 45 users, respectively. We observed that Out of Site substantially changed users' web experiences and that some users preferred to have their actions automated while others simply wanted assistance with awareness of relevant information. We also observed some attempts at non-conformance with respect to boycott goals. Our results support the strong potential of Out of Site and boycott-assisting technologies more generally and inform means by which boycott-assisting technologies can meet this potential.

## ACKNOWLEDGEMENTS

We would like to thank our anonymous reviewers, Nick Vincent, Loren Terveen, and the members of the PSA Research Group (Northwestern University) and GroupLens (University of Minnesota) for all of their helpful feedback on this research. This research was supported in part by Northwestern's Technology and Social Behavior (TSB) program.

## REFERENCES

[1] Veronika A. Andorfer and Ulf Liebe. 2012. Research on Fair Trade Consumption—A Review. *J. Bus. Ethics* 106, 4 (April 2012), 415–435. DOI:https://doi.org/10.1007/s10551-011-1008-5

[2] Lucy Atkinson. 2012. Buying In to Social Change: How Private Consumption Choices Engender Concern for the Collective. *Ann. Am. Acad. Pol. Soc. Sci.* 644, 1 (November 2012), 191–206. DOI:https://doi.org/10.1177/0002716212448366

[3] Clive Barnett, Paul Cloke, Nick Clarke, and Alice Malpass. Consuming Ethics: Articulating the Subjects and Spaces of Ethical Consumption. *Antipode* 37, 1 , 23–45. DOI:https://doi.org/10.1111/j.0066-4812.2005.00472.x

[4] Rajesh Basak, Niloy Ganguly, Shamik Sural, and Soumya K. Ghosh. 2016. Look Before You Shame: A Study on Shaming Activities on Twitter. In *Proceedings of the 25th International Conference Companion on World Wide Web* (WWW '16 Companion), 11–12. DOI:https://doi.org/10.1145/2872518.2889414

[5] Amy B. Becker and Lauren Copeland. 2016. Networked publics: How connective social media use facilitates political consumerism among LGBT Americans. *J. Inf. Technol. Polit.* 13, 1 (January 2016), 22–36. DOI:https://doi.org/10.1080/19331681.2015.1131655

[6] W. Lance Bennett and Alexandra Segerberg. 2012. THE LOGIC OF CONNECTIVE ACTION. *Inf. Commun. Soc.* 15, 5 (June 2012), 739–768. DOI:https://doi.org/10.1080/1369118X.2012.670661

[7] Elizabeth Goodman Ph D. School of Information University of California Berkeley, Mike Kuniavsky, and Andrea Moed. 2012. *Observing the User Experience, Second Edition: A Practitioner's Guide to User Research* (2 edition ed.). Morgan Kaufmann, Amsterdam ; Boston.

[8] Hugh Beyer and Karen Holtzblatt. 1999. Contextual Design. *interactions* 6, 1 (January 1999), 32–42. DOI:https://doi.org/10.1145/291224.291229

[9] Rogue 🔥 Bigham. 2018. it's easy to think that computer science will only have positive broader impacts, or that how it's used isn't your problem: neither is true. my @acm_fca colleagues and I think we should treat Negative Impacts seriously when writing papers & seeking fundinghttps://acm-fca.org/2018/03/29/negativeimpacts/ …. @jeffbigham. Retrieved April 19, 2018 from https://twitter.com/jeffbigham/status/979394988772556800

[10] Emma Boulstridge and Marylyn Carrigan. 2000. Do consumers really care about corporate responsibility? Highlighting the attitude—behaviour gap. *J. Commun. Manag.* 4, 4 (April 2000), 355–368. DOI:https://doi.org/10.1108/eb023532

[11] Lia Bozarth and Ceren Budak. 2017. *Social Movement Organizations in Online Movements*. Social Science Research Network, Rochester, NY. Retrieved January 19, 2018 from https://papers.ssrn.com/abstract=3068546

[12] Marylyn Carrigan and Ahmad Attalla. 2001. The myth of the ethical consumer – do ethics matter in purchase behaviour? *J. Consum. Mark.* 18, 7 (December 2001), 560–578. DOI:https://doi.org/10.1108/07363760110410263

[13] Justin Cheng and Michael Bernstein. 2014. Catalyst: triggering collective action with thresholds. 1211–1221. DOI:https://doi.org/10.1145/2531602.2531635

[14] Mary L. Cummings. 2006. Integrating ethics in design through the value-sensitive design approach. *Sci. Eng. Ethics* 12, 4 (December 2006), 701–715. DOI:https://doi.org/10.1007/s11948-006-0065-0

[15] Munmun De Choudhury, Shagun Jhaver, Benjamin Sugar, and Ingmar Weber. 2016. Social Media Participation in an Activist Movement for Racial Equality. In *Tenth International AAAI Conference on Web and Social Media*.

[16] Tawanna R. Dillahunt, Christopher A. Brooks, and Samarth Gulati. 2015. Detecting and Visualizing Filter Bubbles in Google and Bing. In *Proceedings of the 33rd Annual ACM Conference Extended Abstracts on Human Factors in Computing Systems* (CHI EA '15), 1851–1856. DOI:https://doi.org/10.1145/2702613.2732850

[17] Jill P. Dimond, Michaelanne Dye, Daphne LaRose, and Amy S. Bruckman. 2013. Hollaback!: the role of storytelling online in a social movement organization. In *Proceedings of the 2013 conference on Computer supported cooperative work*, 477–490. DOI:https://doi.org/ 10.1145/2441776.2441831

[18] Tilman Dingler. 2016. Cognition-aware Systems As Mobile Personal Assistants. In *Proceedings of the 2016 ACM International Joint Conference on Pervasive and Ubiquitous Computing: Adjunct* (UbiComp '16), 1035–1040. DOI:https://doi.org/10.1145/2968219.2968565

[19] Daniel A. Epstein, Bradley H. Jacobson, Elizabeth Bales, David W. McDonald, and Sean A. Munson. 2015. From "Nobody Cares" to "Way to Go!": A Design Framework for Social Sharing in Personal Informatics. In *Proceedings of the 18th ACM Conference on Computer Supported Cooperative Work & Social Computing* (CSCW '15), 1622–1636. DOI:https://doi.org/10.1145/2675133.2675135

[20] Robert Epstein, Ronald E. Robertson, David Lazer, and Christo Wilson. 2017. Suppressing the Search Engine Manipulation Effect (SEME). *Proc. ACM Hum.-Comput. Interact.* 1, CSCW (December 2017), 1–22. DOI:https://doi.org/10.1145/3134677

[21] Claudia I. Flores-Saviaga, Brian C. Keegan, and Saiph Savage. 2018. Mobilizing the Trump Train: Understanding Collective Action in a Political Trolling Community. In *Twelfth International AAAI Conference on Web and Social Media*.

[22] Christopher Frauenberger, Amy S. Bruckman, Cosmin Munteanu, Melissa Densmore, and Jenny Waycott. 2017. Research Ethics in HCI: A Town Hall Meeting. In *Proceedings of the 2017 CHI Conference Extended Abstracts on Human Factors in Computing Systems* (CHI EA '17), 1295–1299. DOI:https://doi.org/10.1145/3027063.3051135

[23] Christopher Frauenberger, Marjo Rauhala, and Geraldine Fitzpatrick. 2017. In-Action Ethics. *Interact. Comput.* 29, 2 (March 2017), 220–236. DOI:https://doi.org/10.1093/iwc/iww024

[24] B. Friedman, D. C. Howe, and E. Felten. 2002. Informed consent in the Mozilla browser: implementing value-sensitive design. In *Proceedings of the 35th Annual Hawaii International Conference on System Sciences*, 10 pp.-. DOI:https://doi.org/10.1109/HICSS.2002.994366

[25] Batya Friedman. 1997. *Human Values and the Design of Computer Technology*. Cambridge University Press.

[26] Batya Friedman and Peter H. Kahn Jr. 2003. Human values, ethics, and design. In *The Human-computer Interaction Handbook*, Julie A. Jacko and Andrew Sears (eds.). L. Erlbaum Associates Inc., Hillsdale, NJ, USA, 1177–1201.

[27] Monroe Friedman. 2002. *Consumer Boycotts: Effecting Change Through the Marketplace and Media*. Routledge.

[28] Brayden G King. 2011. The Tactical Disruptiveness of Social Movements: Sources of Market and Mediated Disruption in Corporate Boycotts. *Soc. Probl.* 58, (November 2011), 491–517. DOI:https://doi.org/10.1525/sp.2011.58.4.491

[29] Mark Granovetter. 1978. Threshold Models of Collective Behavior. *Am. J. Sociol.* 83, 6 (May 1978), 1420–1443. DOI:https://doi.org/10.1086/226707

[30] Erin Griffith. 2017. Techies Still Think They're the Good Guys. They're Not. | Backchannel. *Wired.* Retrieved August 27, 2018 from https://www.wired.com/story/the-other-tech-bubble/

[31] Bruce Hanington and Bella Martin. 2012. *Universal Methods of Design: 100 Ways to Research Complex Problems, Develop Innovative Ideas, and Design Effective Solutions* (58480th edition ed.). Rockport Publishers, Beverly, MA.

[32] Richard A. Hawkins. 2010. Boycotts, buycotts and consumer activism in a global context: An overview. *Manag. Organ. Hist.* 5, 2 (May 2010), 123–143. DOI:https://doi.org/10.1177/1744935910361644
[33] Gillian R. Hayes. 2014. Knowing by Doing: Action Research as an Approach to HCI. In *Ways of Knowing in HCI*, Judith S. Olson and Wendy A. Kellogg (eds.). Springer New York, New York, NY, 49–68. DOI:https://doi.org/10.1007/978-1-4939-0378-8_3
[34] Brent Hecht, Lauren Wilcox, Jeffrey Bigham, Johannes Schöning, Ehsan Hoque, Jason Ernst, Yonatan Bisk, Luigi De Russis, Lana Yarosh, Bushra Anjum, Danish Contractor, and Cathy Wu. 2018. It's Time to Do Something: Mitigating the Negative Impacts of Computing Through a Change to the Peer Review Process. *ACM FCA*. Retrieved April 19, 2018 from https://acm-fca.org/2018/03/29/negativeimpacts/
[35] Gabriel Emile Hine, Jeremiah Onaolapo, Emiliano De Cristofaro, Nicolas Kourtellis, Ilias Leontiadis, Riginos Samaras, Gianluca Stringhini, and Jeremy Blackburn. 2017. Kek, Cucks, and God Emperor Trump: A Measurement Study of 4chan's Politically Incorrect Forum and Its Effects on the Web. In *Eleventh International AAAI Conference on Web and Social Media*.
[36] Frank Den Hond and Frank G. A. De Bakker. 2007. Ideologically Motivated Activism: How Activist Groups Influence Corporate Social Change Activities. *Acad. Manage. Rev.* 32, 3 (2007), 901–924. DOI:https://doi.org/10.2307/20159341
[37] Daniel C. Howe and Helen Nissenbaum. 2017. Engineering privacy and protest: A case study of AdNauseam. *CEUR Workshop Proc.* 1873, (2017), 57–64.
[38] Tiffany Hsu. 2018. Big and Small, N.R.A. Boycott Efforts Come Together in Gun Debate. *The New York Times*. Retrieved April 19, 2018 from https://www.nytimes.com/2018/02/27/business/nra-boycotts.html
[39] Lilly C. Irani and M. Six Silberman. 2013. Turkopticon: Interrupting Worker Invisibility in Amazon Mechanical Turk. In *Proceedings of the SIGCHI Conference on Human Factors in Computing Systems* (CHI '13), 611–620. DOI:https://doi.org/10.1145/2470654.2470742
[40] Mike Isaac. 2018. What You Need to Know About #DeleteUber. *The New York Times*. Retrieved July 5, 2018 from https://www.nytimes.com/2017/01/31/business/delete-uber.html
[41] Shagun Jhaver, Larry Chan, and Amy Bruckman. 2018. The view from the other side: The border between controversial speech and harassment on Kotaku in Action. *First Monday* 23, 2 (February 2018). DOI:https://doi.org/10.5210/fm.v23i2.8232
[42] Yea-Seul Kim, Jessica Hullman, and Maneesh Agrawala. 2016. Generating Personalized Spatial Analogies for Distances and Areas. In *Proceedings of the 2016 CHI Conference on Human Factors in Computing Systems* (CHI '16), 38–48. DOI:https://doi.org/10.1145/2858036.2858440
[43] Jill Gabrielle Klein, N. Craig Smith, and Andrew John. 2004. Why We Boycott: Consumer Motivations for Boycott Participation. *J. Mark.* 68, 3 (2004), 92–109.
[44] Sebastian Koos. 2012. What drives political consumption in Europe? A multi-level analysis on individual characteristics, opportunity structures and globalization. *Acta Sociol.* 55, 1 (March 2012), 37–57. DOI:https://doi.org/10.1177/0001699311431594
[45] Adam D. I. Kramer, Jamie E. Guillory, and Jeffrey T. Hancock. 2014. Experimental evidence of massive-scale emotional contagion through social networks. *Proc. Natl. Acad. Sci.* (May 2014), 201320040. DOI:https://doi.org/10.1073/pnas.1320040111
[46] Robert E. Kraut, Paul Resnick, Sara Kiesler, Moira Burke, Yan Chen, Niki Kittur, Joseph Konstan, Yuqing Ren, and John Riedl. 2012. *Building Successful Online Communities: Evidence-Based Social Design*. MIT Press.
[47] Christopher A. Le Dantec, Erika Shehan Poole, and Susan P. Wyche. 2009. Values As Lived Experience: Evolving Value Sensitive Design in Support of Value Discovery. In *Proceedings of the SIGCHI Conference on Human Factors in Computing Systems* (CHI '09), 1141–1150. DOI:https://doi.org/10.1145/1518701.1518875
[48] Kimberly Ling, Gerard Beenen, Pamela Ludford, Xiaoqing Wang, Klarissa Chang, Xin Li, Dan Cosley, Dan Frankowski, Loren Terveen, Al Mamunur Rashid, Paul Resnick, and Robert Kraut. 2005. Using Social Psychology to Motivate Contributions to Online Communities. *J. Comput.-Mediat. Commun.* 10, 4 (July 2005), 00–00. DOI:https://doi.org/10.1111/j.1083-6101.2005.tb00273.x
[49] Tetyana Lokot and Nicholas Diakopoulos. 2016. News Bots: Automating news and information dissemination on Twitter. *Digit. Journal.* 4, 6 (August 2016), 682–699. DOI:https://doi.org/10.1080/21670811.2015.1081822
[50] Chun-Ta Lu, Sihong Xie, Xiangnan Kong, and Philip S. Yu. 2014. Inferring the Impacts of Social Media on Crowdfunding. In *Proceedings of the 7th ACM International Conference on Web Search and Data Mining* (WSDM '14), 573–582. DOI:https://doi.org/10.1145/2556195.2556251
[51] J. Nathan Matias. 2016. Going Dark: Social Factors in Collective Action Against Platform Operators in the Reddit Blackout. In *Proceedings of the 2016 CHI Conference on Human Factors in Computing Systems* (CHI '16), 1138–1151. DOI:https://doi.org/10.1145/2858036.2858391

[52] Connor McMahon, Isaac Johnson, and Brent Hecht. 2017. The Substantial Interdependence of Wikipedia and Google: A Case Study on the Relationship Between Peer Production Communities and Information Technologies. In *Eleventh International AAAI Conference on Web and Social Media*.

[53] Michele Micheletti, Andreas Follesdal, and Dietlind Stolle. Politics, Products, and Markets: Exploring Political Consumerism Past and Present. *Econ. Geogr.* 84, 1 , 123–125. DOI:https://doi.org/10.1111/j.1944-8287.2008.tb00400.x

[54] Michael J. Muller. 2003. Participatory design: the third space in HCI. In *The Human-computer Interaction Handbook*, Julie A. Jacko and Andrew Sears (eds.). L. Erlbaum Associates Inc., Hillsdale, NJ, USA, 1051–1068.

[55] Sean A. Munson, Stephanie Y. Lee, and Paul Resnick. 2013. Encouraging Reading of Diverse Political Viewpoints with a Browser Widget. In *Proceedings of the 7th International Conference on Weblogs and Social Media, ICWSM 2013*.

[56] Terry Newholm. 2005. Case Studying Ethical Consumers' Projects and Strategies. In *The Ethical Consumer*. SAGE Publications Ltd, London, 107–124. DOI:https://doi.org/10.4135/9781446211991

[57] Terry Newholm and Deirdre Shaw. 2007. Studying the ethical consumer: a review of research. *J. Consum. Behav.* 6, 5 (October 2007), 253–270. DOI:https://doi.org/10.1002/cb.225

[58] Benjamin J. Newman and Brandon L. Bartels. 2011. Politics at the Checkout Line: Explaining Political Consumerism in the United States. *Polit. Res. Q.* 64, 4 (2011), 803–817.

[59] Tien T. Nguyen, Pik-Mai Hui, F. Maxwell Harper, Loren Terveen, and Joseph A. Konstan. 2014. Exploring the Filter Bubble: The Effect of Using Recommender Systems on Content Diversity. In *Proceedings of the 23rd International Conference on World Wide Web* (WWW '14), 677–686. DOI:https://doi.org/10.1145/2566486.2568012

[60] Jakob Nielsen. 1994. Guerrilla HCI: using discount usability engineering to penetrate the intimidation barrier. In *Cost-justifying usability*, Randolph G. Bias and Deborah J. Mayhew (eds.). Academic Press, Inc., Orlando, FL, USA, 245–272.

[61] Jakob Nielsen. 1994. *Usability Engineering*. Elsevier.

[62] Jakob Nielsen and Rolf Molich. 1990. Heuristic Evaluation of User Interfaces. In *Proceedings of the SIGCHI Conference on Human Factors in Computing Systems* (CHI '90), 249–256. DOI:https://doi.org/10.1145/97243.97281

[63] Mancur OLSON. 2009. *The Logic of Collective Action: Public Goods and the Theory of Groups, Second printing with new preface and appendix*. Harvard University Press.

[64] Tapan Parikh. 2018. Mitigating the Negative Implications of Computing: Making Space for Debate. *Medium*. Retrieved April 19, 2018 from https://medium.com/@tap2k/mitigating-the-negative-implications-of-computing-making-space-for-debate-b04410f3a82b

[65] Eli Pariser. 2012. *The Filter Bubble: How the New Personalized Web is Changing what We Read and how We Think*. Penguin Books.

[66] Jenny Preece, Helen Sharp, and Yvonne Rogers. 2015. *Interaction Design: Beyond Human-Computer Interaction* (4 edition ed.). Wiley, Chichester.

[67] Al M. Rashid, Kimberly Ling, Regina D. Tassone, Paul Resnick, Robert Kraut, and John Riedl. 2006. Motivating Participation by Displaying the Value of Contribution. In *Proceedings of the SIGCHI Conference on Human Factors in Computing Systems* (CHI '06), 955–958. DOI:https://doi.org/10.1145/1124772.1124915

[68] Paul Resnick, R. Kelly Garrett, Travis Kriplean, Sean A. Munson, and Natalie Jomini Stroud. 2013. Bursting Your (Filter) Bubble: Strategies for Promoting Diverse Exposure. In *Proceedings of the 2013 Conference on Computer Supported Cooperative Work Companion* (CSCW '13), 95–100. DOI:https://doi.org/10.1145/2441955.2441981

[69] Daniel E. Rose and Danny Levinson. 2004. Understanding User Goals in Web Search. In *Proceedings of the 13th International Conference on World Wide Web* (WWW '04), 13–19. DOI:https://doi.org/10.1145/988672.988675

[70] Saiph Savage, Andres Monroy-Hernandez, and Tobias Höllerer. 2016. Botivist: Calling Volunteers to Action Using Online Bots. In *Proceedings of the 19th ACM Conference on Computer-Supported Cooperative Work & Social Computing* (CSCW '16), 813–822. DOI:https://doi.org/10.1145/2818048.2819985

[71] Samantha Schmidt. 2017. Sean Hannity's fans call for Keurig boycott after coffeemaker company pulls ads from his show. *Washington Post*. Retrieved April 19, 2018 from https://www.washingtonpost.com/news/morning-mix/wp/2017/11/13/sean-hannitys-fans-call-for-keurig-boycott-after-coffee-maker-pulls-ads-from-his-show/

[72] Paul Sergius Koku. 2012. On the effectiveness of consumer boycotts organized through the internet: the market model. *J. Serv. Mark.* 26, 1 (February 2012), 20–26. DOI:https://doi.org/10.1108/08876041211199698

[73] Aaron Shaw, Haoqi Zhang, Andrés Monroy-Hernández, Sean Munson, Benjamin Mako Hill, Elizabeth Gerber, Peter Kinnaird, and Patrick Minder. 2014. Computer Supported Collective Action. *interactions* 21, 2 (March 2014), 74–77. DOI:https://doi.org/10.1145/2576875

[74] Kate Starbird. 2013. Delivering Patients to Sacré Coeur: Collective Intelligence in Digital Volunteer Communities. In *Proceedings of the SIGCHI Conference on Human Factors in Computing Systems* (CHI '13), 801–810. DOI:https://doi.org/10.1145/2470654.2470769

[75] Kate Starbird, Ahmer Arif, Tom Wilson, Katherine Van Koevering, Katya Yefimova, and Daniel Scarnecchia. 2018. Ecosystem or Echo-System? Exploring Content Sharing across Alternative Media Domains. In *Twelfth International AAAI Conference on Web and Social Media*.

[76] Dietlind Stolle, Marc Hooghe, and Michele Micheletti. 2005. Politics in the Supermarket: Political Consumerism as a Form of Political Participation. *Int. Polit. Sci. Rev.* 26, 3 (July 2005), 245–269. DOI:https://doi.org/10.1177/0192512105053784

[77] Marshall Van Alstyne and Erik Brynjolfsson. 2005. Global Village or Cyber-Balkans? Modeling and Measuring the Integration of Electronic Communities. *Manag. Sci.* 51, 6 (June 2005), 851–868. DOI:https://doi.org/10.1287/mnsc.1050.0363

[78] Moshe Y. Vardi. 2018. Move Fast and Break Things. *Communications of the ACM 61*, 7.

[79] Ion Bogdan Vasi and Michael Macy. 2002. The Mobilizer's Dilemma: Crisis, Empowerment, and Collective Action. *Soc. Forces* 81, (2003 2002), 979–998.

[80] Haiyi Zhu, Amy Zhang, Jiping He, Robert E. Kraut, and Aniket Kittur. 2013. Effects of Peer Feedback on Contribution: A Field Experiment in Wikipedia. In *Proceedings of the SIGCHI Conference on Human Factors in Computing Systems* (CHI '13), 2253–2262. DOI:https://doi.org/10.1145/2470654.2481311

[81] Homero Gil de Zúñiga, Lauren Copeland, and Bruce Bimber. 2014. Political consumerism: Civic engagement and the social media connection. *New Media Soc.* 16, 3 (May 2014), 488–506. DOI:https://doi.org/10.1177/1461444813487960

[82] 2017. Standing up for what's right | Uber Newsroom US. *Uber Newsroom*. Retrieved August 27, 2018 from https://www.uber.com/newsroom/standing-up-for-whats-right-3/

[83] 2018. Deep Fakes: A Looming Crisis for National Security, Democracy and Privacy? *Lawfare*. Retrieved April 19, 2018 from https://www.lawfareblog.com/deep-fakes-looming-crisis-national-security-democracy-and-privacy

[84] 2018. AI at Google: our principles. *Google*. Retrieved August 27, 2018 from https://www.blog.google/technology/ai/ai-principles/

[85] 2018. New Toolkit Shows Companies How to Anticipate, Prevent Bad Actors from Using Tech in Harmful Ways. *The Institute for the Future and Omidyar Network*. Retrieved from https://ethicalos.org/wp-content/uploads/2018/08/IFTF_Ethical-OS-press-release_8.7.18.pdf

[86] #GrabYourWallet adds Chrome extension to make avoiding Trump brands even easier. Retrieved April 19, 2018 from https://mic.com/articles/171396/grab-your-wallet-adds-chrome-extension-to-make-avoiding-trump-brands-even-easier

[87] The woman behind the boycott that is pressuring retailers to dump the Trumps. *Washington Post*. Retrieved August 26, 2018 from https://www.washingtonpost.com/news/business/wp/2017/02/13/the-woman-behind-the-boycott-that-is-pressuring-retailers-to-dump-the-trumps/

[88] #GrabYourWallet's Anti-Trump Boycott Looks To Expand Its Reach. *NPR.org*. Retrieved August 26, 2018 from https://www.npr.org/2017/04/16/523960521/-grabyourwallets-anti-trump-boycott-looks-to-expand-its-reach

[89] Buycott. *Buycott App*. Retrieved July 11, 2018 from https://www.buycott.com/

[90] ACM Ethics. *ACM Ethics*. Retrieved July 11, 2018 from https://ethics.acm.org/

[91] Groups. *Southern Poverty Law Center*. Retrieved April 19, 2018 from https://www.splcenter.org/fighting-hate/extremist-files/groups

[92] Types of Discrimination. Retrieved April 19, 2018 from https://www.eeoc.gov/laws/types/

[93] App Store Review Guidelines - Apple Developer. Retrieved April 19, 2018 from https://developer.apple.com/app-store/review/guidelines/#objectionable-content

[94] Resistbot. Retrieved July 11, 2018 from https://resist.bot